\documentclass{article}
\usepackage[T1]{fontenc}
\usepackage[utf8]{inputenc}
\usepackage[]{ismir} 
\usepackage{amsmath,cite,url}
\usepackage{graphicx}
\usepackage{color}
\usepackage{booktabs}
\usepackage{multirow}
\usepackage{array}
\usepackage{hhline}
\usepackage{colortbl}

\usepackage{tabularray}
\usepackage{xcolor}
\usepackage{booktabs}
\usepackage{tabularx}
\usepackage{makecell}
\usepackage{pifont}
\newcommand{\cmark}{\ding{51}} 
\newcommand{\xmark}{\ding{55}}

\title{PianoBind: A Multimodal Joint Embedding Model for Pop-piano Music}

\multauthor
{Hayeon Bang \hspace{1cm} Eunjin Choi \hspace{1cm} Seungheon Doh \hspace{1cm} {\bf Juhan Nam}}{
Graduate School of Culture Technology, KAIST, South Korea\\
{\tt\small \{hayeonbang,jech,seungheondoh,juhan.nam\}@kaist.ac.kr}
}

\def\authorname{Hayeon Bang, Eunjin Choi, Seungheon Doh, and Juhan Nam}

\begin{document}

\maketitle

\begin{abstract}
Solo piano music, despite being a single-instrument medium, possesses significant expressive capabilities, conveying rich semantic information across genres, moods, and styles. However, current general-purpose music representation models, predominantly trained on large-scale datasets, often struggle to capture subtle semantic distinctions within homogeneous solo piano music. Furthermore, existing piano-specific representation models are typically unimodal, failing to capture the inherently multimodal nature of piano music, expressed through audio, symbolic, and textual modalities. To address these limitations, we propose~\textbf{PianoBind}, a piano-specific multimodal joint embedding model. We systematically investigate strategies for multi-source training and modality utilization within a joint embedding framework optimized for capturing fine-grained semantic distinctions in (1) small-scale and (2) homogeneous piano datasets. Our experimental results demonstrate that PianoBind learns multimodal representations that effectively capture subtle nuances of piano music, achieving superior text-to-music retrieval performance on in-domain and out-of-domain piano datasets compared to general-purpose music joint embedding models. Moreover, our design choices offer reusable insights for multimodal representation learning with homogeneous datasets beyond piano music.
\end{abstract}

\section{Introduction}\label{sec:introduction}

The piano stands as a uniquely versatile solo instrument capable of conveying complex polyphonic musical expression through a single instrument. With its expansive tonal range, harmonic possibilities, and expressive capabilities—even allowing orchestral works to be effectively performed on a single keyboard—piano music encompasses diverse genres, styles, and expressive content. 
Numerous studies in the Music Information Retrieval (MIR) field have targeted tasks focused on piano music, such as piano music generation \cite{popmusic, compound, gen2, gen3} and automatic music transcription \cite{amt1, amt2, amt3, amt4}. However, research on piano-specific representation models remains limited. Existing approaches are typically constrained to a single modality~\cite{midibert, pianobart}, failing to reflect the inherently multimodal nature of piano music that encompasses audio recordings, symbolic MIDI, and semantic descriptions.


Recent advances in multimodal joint embedding models have shown promise in bridging the gap between audio and text domains \cite{mulan, muscall, clap, ttmr, ttmrpp} and symbolic domains \cite{clamp, clamp2, clamp3}, yet they often fall short in specialized domains—particularly in solo piano music. Despite their versatility across diverse music categories, general-purpose models typically lack the sensitivity needed to capture subtle semantic nuances within homogeneous solo piano music. This limitation arises primarily from the scarcity of high-quality piano-specific data in general-purpose music-text datasets, which are typically dominated by multi-instrumental or vocal-centric music. 


\begin{figure}
  \centering
  \includegraphics[alt={ISMIR 2025 template example image},width=1.0\linewidth]{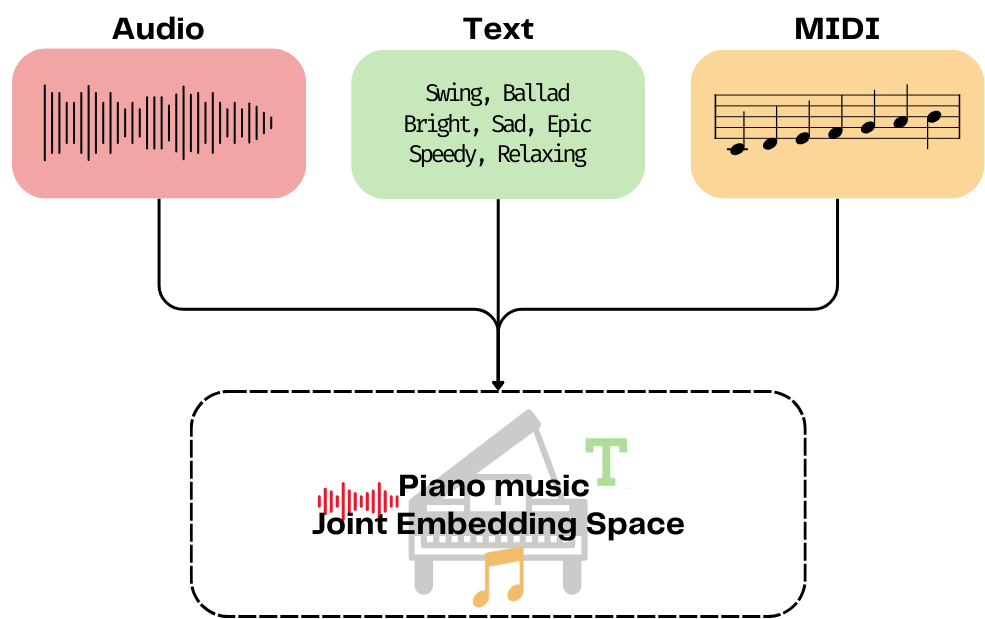}
  \caption{Illustration of PianoBind: A multimodal piano music representation model integrating audio, MIDI, and text.}
    \vspace{-10pt}
  \label{fig:figure1}
\end{figure}

In this work, we present PianoBind, a multimodal joint embedding model that integrates multiple modalities of solo piano music—audio, symbolic (MIDI), and textual descriptions—within a unified embedding space (illustrated in Figure \ref{fig:figure1}), enabling a more comprehensive representation. Our goal is to capture the fine-grained semantic characteristics of piano music—spanning genre, mood, and style—that are often overlooked by large-scale, general-purpose models. To achieve this, we use the PIAST dataset \cite{piast} for training, a pop-piano music (e.g. new-age, piano cover, jazz and its sub-genres) dataset that includes audio, MIDI, and textual descriptions. We systematically explore multi-source training strategies tailored to the characteristics of domain-specific datasets. We utilize a comparatively large amount of automatically collected textual data with weaker temporal alignment to audio. This approach compensates for the limited amount of human-annotated data. Furthermore, we propose effective methods for combining multimodal information both during training and at retrieval time, where joint audio-symbolic embeddings can significantly enhance text-to-music retrieval, particularly in distinguishing highly similar piano pieces.

Our experiments on both in-domain and out-of-domain piano datasets show that PianoBind outperforms general-purpose music joint embedding models in capturing the nuances of piano solo music. By focusing on small-scale, homogeneous datasets, our findings also offer valuable guidelines for developing specialized multimodal representation learning approaches in other domains with limited data. As such, this study contributes to the growing body of piano-centric MIR research. It also contributes to broader discussions on efficient and fine-grained multimodal modeling. These contributions are especially relevant when data availability is inherently constrained. We have publicly released code and pretrained weights of PianoBind, with the demo online\footnote{\url{https://hayeonbang.github.io/PianoBind/}}.

\section{Related Works}
\subsection{Piano Music Representation Learning}

Piano music has long served as a central subject in MIR, owing to its structural richness and expressive depth. However, despite this sustained attention, existing representation learning approaches for piano music are predominantly unimodal—relying solely on symbolic or audio data—thus failing to reflect the inherently multimodal nature of piano music.

Early piano music representation in the symbolic music domain primarily focused on disentangling low-level symbolic features, mostly with the goal of enhancing controllability in music generation tasks. Models such as PianoTree VAE~\cite{pianotree}, Wang et al.~\cite{polydis}, and CollageNet~\cite{collagenet} exemplify this approach, decomposing music into attributes such as rhythm, harmony, texture, and structure to facilitate user-controlled music generation. 

More recently, transformer-based models like MidiBERT-Piano~\cite{midibert} and PianoBART~\cite{pianobart} have expanded piano music understanding through large-scale pretraining, capturing both low-level features and higher-level musical attributes. MidiBERT-Piano introduced masked modeling objectives for solo piano MIDI data, demonstrating strong transferability to downstream tasks such as composer classification and expressive attribute prediction. Building upon this foundation, PianoBART extended these capabilities from understanding toward generation tasks, facilitating more sophisticated music creation and symbolic inpainting tasks. Nonetheless, these models remain purely symbolic, lacking integration with acoustic information or natural language semantics.


Despite the progress in piano music representation learning, multimodal understanding of piano music has been limited, primarily due to the scarcity of datasets that support such approaches. Until recently, few resources existed that combined multiple modalities of piano performance. EMOPIA~\cite{emopia} made an initial step by providing paired audio and MIDI recordings with emotion labels, while PIAST~\cite{piast} has recently extended this further by adding comprehensive textual annotations describing genre and mood. However, even with these multimodal datasets becoming available, no prior research has proposed an integrated approach that jointly leverages audio, symbolic, and textual modalities for piano music understanding.


\subsection{Multimodal Joint Embedding Models}



Multimodal joint embedding models aim to align data from different modalities in a shared embedding space. This alignment creates semantically meaningful representations that capture relationships between modalities, enabling cross-modal retrieval, understanding, and conditioned generation. MuLan~\cite{mulan} pioneered large-scale audio–text training with over 44 million pairs, showing strong retrieval capabilities. MusCALL~\cite{muscall} proposed an audio-text dual encoder architecture, demonstrating improved retrieval performance and downstream performance. CLAP~\cite{clap} expanded on this by leveraging audio captioning corpora and keyword-to-caption augmentation strategies, and introduced a joint audio-text embedding model. Further refinements include models like TTMR~\cite{ttmr} and TTMR++~\cite{ttmrpp}, which address varying query granularities—ranging from single tags to full sentences—and incorporate rich metadata to produce more descriptive and contextual text embeddings. These methods improved retrieval accuracy by modeling both linguistic and musical subtleties.

In parallel, symbolic-text joint embedding models have been explored through the CLaMP series~\cite{clamp, clamp2, clamp3}, which introduced contrastive learning between symbolic music (e.g., ABC or MIDI) and textual descriptions. While CLaMP pioneered symbolic music retrieval through text-ABC joint training, CLaMP2 expanded this to multilingual text and MIDI data, and CLaMP3 further incorporated audio and performance signals, marking the first trimodal framework in music representation learning to jointly align audio, symbolic, and textual modalities. 
\begin{figure*}[hbt!]
  \centering
  \includegraphics[alt={The overall training strategies of PianoBind: Multi-source training, Trimodal learning, and Multimodal Item Embedding.},width=0.86\linewidth]{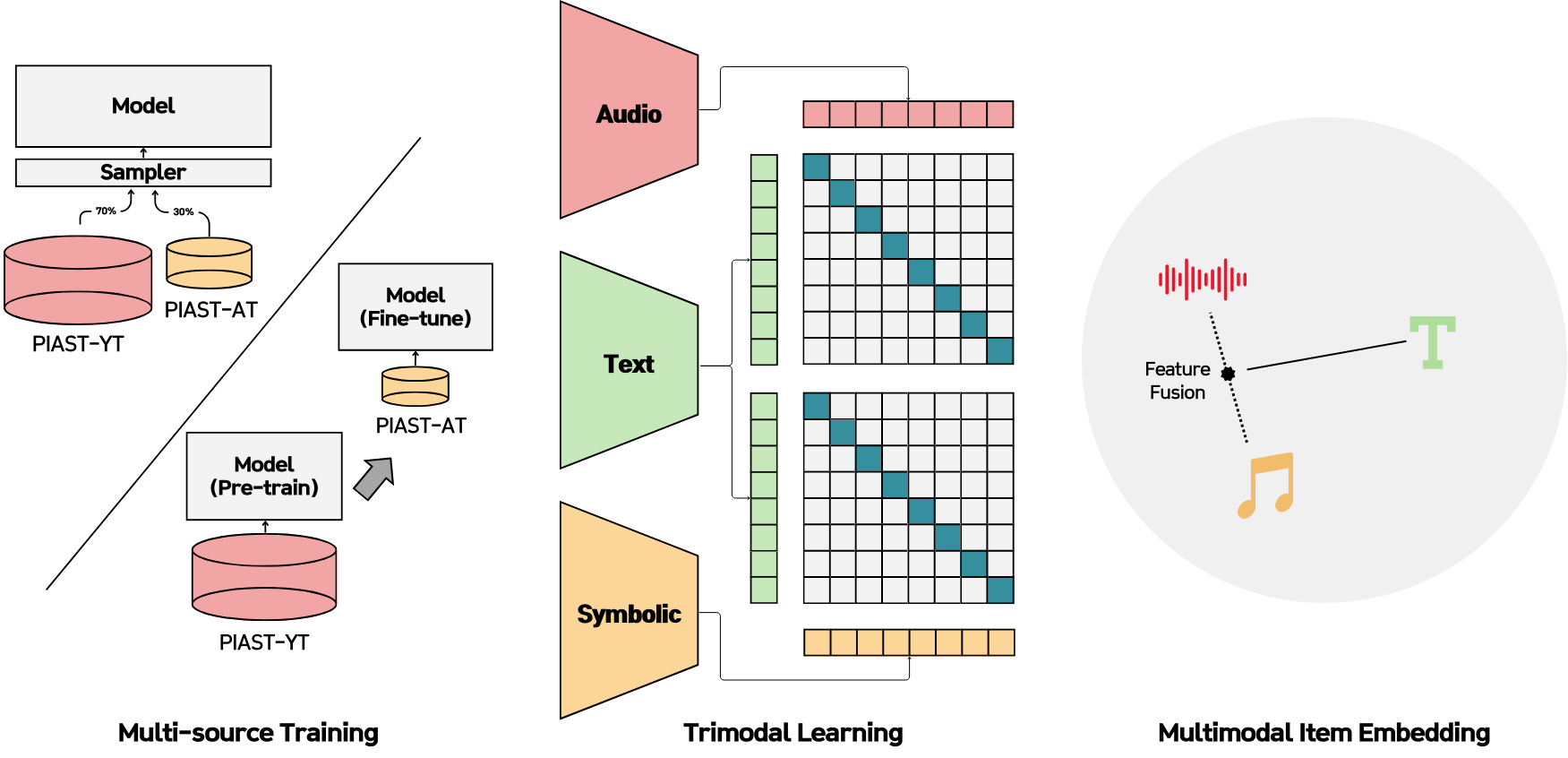}
    \caption{Training strategies of PianoBind: (1) Multi-source training combines small strongly aligned (human-annotated) and large weakly aligned (automatically collected) datasets; (2) Trimodal learning simultaneously aligns audio, symbolic, and textual embeddings; (3) Multimodal item embedding merges multiple modalities into a unified representation.}
    \vspace{-9pt}
  \label{fig:model}
\end{figure*}

Despite these advances, most current models are trained on general-purpose, large-scale datasets covering a broad range of musical styles and instrumentation. As a result, they often fall short of capturing the fine-grained semantic differences required for more homogeneous domains like solo piano music. These models tend to underperform in settings where subtle variations in genre, mood, and style must be accurately distinguished.

To address this gap, our proposed model PianoBind integrates piano-specific audio, symbolic, and textual modalities in a unified embedding space, enabling more precise retrieval within this specialized musical domain.

\section{PianoBind}\label{sec:model}
Considering the specific characteristics of piano solo datasets—such as homogeneous data distribution, limited dataset size, and multimodality—we propose a multimodal joint embedding model specialized for solo piano music. In this section, we describe the overall architecture of PianoBind (section \ref{sec:overview}), and the training strategies it explores (section \ref{section:framework}), comprising: (1) multi-source learning with strongly and weakly aligned pairs, and (2) modality integration across audio, symbolic, and textual features.

\subsection{Architecture Overview}{\label{sec:overview}}
\subsubsection{Audio Encoder}
Following previous work~\cite{muscall}, we adopt a modified ResNet-50~\cite{resnet} architecture to process mel-spectrogram representations of piano recordings. We extract 128-band mel-spectrograms with a 1024-point FFT, 512-point hop length, and apply log-scaling. As in MusCALL, we apply three stem convolutional layers followed by average pooling, and implement anti-aliased blur pooling. We also follow the downsizing strategy employed in the previous work.

\subsubsection{Symbolic Encoder}
We utilize MidiBERT~\cite{midibert} as a symbolic encoder. The model handles bar, position, pitch, and duration information of MIDI by adapting Compound Word (CP) representation \cite{compound} into its model. 
Since MidiBERT does not contain the [CLS] token, we obtained the final MIDI embedding by mean-pooling over the sequence of hidden states from the last Transformer layer, resulting in a dense representation that is then projected to the shared embedding space.

\subsubsection{Text Encoder}
Our text processing pipeline employs RoBERTa~\cite{roberta}, using its byte-pair encoding (BPE) tokenizer. The tokenized input passes through 12 Transformer layers with 768-dimensional hidden states. Since RoBERTa lacks the standard pooler output found in BERT, we create sentence-level representations by mean-pooling across the final layer's hidden states. These textual embeddings are also mapped to our shared space through a linear projection layer.

\subsubsection{Joint Embedding with Contrastive Loss}
The representations from our three encoders (audio, MIDI, and text) are aligned through modality-specific linear projections into a 512-dimensional shared embedding space. All embeddings undergo $\ell_2$-normalization to ensure consistent scaling across modalities. Cross-modal similarities are computed via dot products between these normalized embeddings, providing the foundation for our various training objectives. We use the N-pair Contrastive loss, known as the InfoNCE~\cite{infoNCE} loss, which maximizes the cosine similarity between positive music-text embedding pairs while minimizing the similarity for negative pairs. For audio-text alignment, the InfoNCE loss is defined as follows:
\begin{equation}
\mathcal{L}_{a \rightarrow t} = -\frac{1}{N}\sum_{i=1}^{N} \log \frac{\exp(z_{a,i} \cdot z_{t,i}^{+} / \tau)}{\sum_{z \in \{z_{t,i}^{+}, z_{t,i}^{-}\}} \exp(z_{a,i} \cdot z / \tau)}
\end{equation}
where $z_{a,i}$ and $z_{t,i}$ refer to audio and text embeddings respectively, in the audio-text training. An analogous loss is computed for MIDI–text pairs by substituting audio embeddings with symbolic ones. $\tau$ represents the temperature parameter, and $z_{t,i}^{-}$ denotes a set of negative text embeddings. This aims to align embeddings between relevant music-text pairs while separating irrelevant pairs in the embedding space. The total loss is computed by symmetrically combining losses in both directions (music-to-text and text-to-music). The final symmetric loss is defined as:
\begin{equation}
\mathcal{L}{a \leftrightarrow t} = \frac{\mathcal{L}_{a \rightarrow t} + \mathcal{L}_{t \rightarrow a}}{2}\
\end{equation} 

\subsection{Training Framework}\label{section:framework}

\subsubsection{Multi-source Training}
Considering the small size and specialized annotation required for piano datasets, we explore two strategies to effectively leverage both large-scale, weakly aligned audio–text data and small-scale, expert-annotated data.

\vspace{2mm} \noindent \textbf{Combined Training}  
We jointly train the model on both sources by mixing them within training batches following prior work~\cite{mt3, joint, clap, ttmrpp}. To mitigate data imbalance and noise from weak supervision, we carefully control the sampling ratio between the two sources. This enables the model to learn generalizable language–music alignments while gradually incorporating domain-specific subtleties of solo piano music.  

\vspace{2mm} \noindent \textbf{Pre-training and Fine-tuning}  
Alternatively, we also adopt the two-stage training strategy, pre-training and fine-tuning, following~\cite{lpcaps}. The model is first pre-trained on the large-scale, weakly labeled dataset to acquire generalizable music representations. It is then fine-tuned on the smaller, expert-annotated data, with encoder parameters updated. This sequential strategy enables the model to benefit from broad semantic coverage during pre-training, while later adapting to the expressive and stylistic nuances of solo piano music through fine-tuning.


\subsubsection{Trimodal Representation Learning}
To fully exploit the multimodal characteristics of piano music, we extend beyond traditional bimodal setups by integrating audio, symbolic (MIDI), and text modalities into a unified retrieval framework. Inspired by the training strategy from AudioCLIP~\cite{audioclip}, we compute the contrastive loss across modality pairs (audio-text, MIDI-text) and average them to form the final objective. However, considering the nature of our MIDI data as transcribed representations derived directly from corresponding audio recordings, including an audio-MIDI loss would not significantly contribute additional semantic distinction. Consequently, we utilize only the audio-text and MIDI-text contrastive losses, calculating their average as our final training objective:
\begin{equation}
\mathcal{L}_{total} = \frac{\mathcal{L}_{a \leftrightarrow t} + \mathcal{L}_{m \leftrightarrow t}}{2}
\end{equation}

Furthermore, we leverage multimodal information not only during the training phase but also at evaluation time. To leverage the complementary strengths of different modalities, we propose multimodal item embeddings that fuse audio and MIDI information. Specifically, audio and MIDI embeddings are integrated through average fusion during the evaluation process.

\section{Experiment}
\subsection{Dataset} 
This study utilizes the PIAST dataset\cite{piast}, the first music-text dataset explicitly designed for pop-piano music. The dataset consists of audio, MIDI, and textual descriptions, based on a comprehensive piano-specific taxonomy of 31 semantic tags across genre, emotion/mood, and style. The dataset comprises two subsets:
PIAST-YT, a large-scale collection of approximately 7,367 tracks (about 900 hours) automatically collected from YouTube, with accompanying textual metadata (titles, descriptions, and tags) refined using a large language model; and
PIAST-AT, a smaller, expert-annotated set of 1,986 tracks (about 17 hours in total, ~30 seconds per track). For both subsets, MIDI data is generated via automatic piano transcription. The transcribed MIDI files were synchronized to downbeat estimates, and melody and chord information was extracted.

Since the textual data of PIAST-YT is automatically collected, it exhibits weak alignment with the audio content, introducing considerable noise in the text-audio relationships. To address this challenge, we apply the two multi-source learning strategies described in Section~\ref{sec:model}, leveraging both the large-scale but weakly aligned PIAST-YT data and the smaller, high-quality PIAST-AT annotations. This combined approach enables more robust representation learning despite the inherent data limitations. For experiments, we use a 9:1 train–validation split for PIAST-YT, and an 8:1:1 train–validation–test split for PIAST-AT.


%
\definecolor{lightgray}{gray}{0.95} 

\begin{table*}[ht]
\centering
{
\begin{tabular}{lcccccccc}
\toprule
\multirow{2}{*}{\textbf{Training Strategy}} 
& \multicolumn{4}{c}{\textbf{In-domain} (199 tracks)} 
& \multicolumn{4}{c}{\textbf{Out-of-domain} (88 tracks)} \\
\cmidrule(lr){2-5} \cmidrule(lr){6-9}
& R@1 & R@5 & R@10 & MedR↓ 
& R@1 & R@5 & R@10 & MedR↓ \\
\midrule
\rowcolor{lightgray}
\multicolumn{9}{l}{\textit{Combined Training}} \\
Audio     & 8.04  & 25.62 & 37.18 & 17 & 2.56  & 10.26 & 35.90 & 17 \\
Symbolic  & 4.02  & 17.09 & 27.13 & 27 & 2.56  & 28.21 & 43.59 & 13 \\
Trimodal  & 6.53  & 24.12 & 37.68 & 17 & 5.13  & 25.64 & 46.15 & 14 \\
\midrule
\rowcolor{lightgray}
\multicolumn{9}{l}{\textit{Pre-training \& Fine-tuning}} \\
Audio     & 6.53  & 28.14 & 42.71 & 15 & 7.69  & 20.51 & 46.15 & 12 \\
Symbolic  & 8.04  &  26.63   & 45.23 & 12 & 5.13  & 20.51 & 35.90 & 12 \\
\textbf{Trimodal}  
          & \textbf{10.55} & \textbf{35.67} & \textbf{52.76} & \textbf{10} 
          & \textbf{15.38} & \textbf{41.03} & \textbf{51.28} & \textbf{10} \\
\bottomrule
\end{tabular}
}
\caption{Performance comparison on text-based music retrieval tasks, on the in-domain (PIAST-AT) and out-of-domain (EMOPIA-Caps) datasets.}
\vspace{-10pt}
\label{table:table1}
\end{table*}

\subsection{Evaluation}
We evaluate our model using both in-domain and out-of-domain text-to-music retrieval tasks. For in-domain evaluation, we use the 10\% held-out test split from PIAST-AT, comprising 199 tracks. For out-of-domain evaluation, we introduce EMOPIA-Caps by manually annotating descriptive tag labels for the EMOPIA test split~\cite{emopia}, and transforming them into natural language captions using a large language model. The initial tags were naturally overlapping with the vocabulary used in the PIAST-AT piano-music taxonomy. To address this and better approximate real-world natural user queries, we paraphrased the tags into free-form natural language captions using GPT-4o~\cite{gpt}. The generated captions were then reviewed and refined by a human music expert with a major in composition,  to ensure their semantic accuracy and musical relevance. This transformation not only better approximates user-style queries, but also enables evaluation of the model’s ability to generalize to diverse textual expressions. Both evaluation datasets use sentence-form textual inputs; however, the in-domain set consists of concatenated tags, whereas the out-of-domain set contains free-form, natural language captions.

For both evaluation settings, we perform text-to-music retrieval using Recall@K (R@1, R@5, R@10) and Median Rank (MedR), as these metrics reflect the model’s ability to generalize to diverse and unconstrained textual queries.

\subsection{Implementation Details}
For audio processing, we use 20-second signals with a sampling rate of 16 kHz, consistent with methods established by previous work~\cite{muscall}. To ensure temporal alignment between audio and MIDI, we match each audio segment's start time (in seconds) with the nearest MIDI bar onset, extracting MIDI tokens from bars that correspond to each audio segment. These sequences are subsequently standardized to exactly 512 tokens through padding or truncation. For text inputs, we employ the RoBERTa tokenizer~\cite{roberta} with a 77-token length limit. To combat overfitting and enhance the diversity of our textual data, we implement a dynamic text dropout strategy building on approaches from several recent works~\cite{ttmr, clamp, ttmrpp}. This technique randomly selects and combines available textual elements (such as tags and captions) in varying orders for each training instance. For combined training, we use a 7:3 sampling ratio between PIAST-YT and PIAST-AT.

All models are trained using the AdamW optimizer with a 5e-5 initial learning rate, 0.2 weight decay, and a consistent batch size of 64 across all experiments. For the contrastive loss, we jointly optimize the temperature parameter $\tau$ alongside encoder and projection parameters, following successful approaches demonstrated in recent multimodal works~\cite{muscall, ttmr}. We select optimal model checkpoints based on median rank performance on our validation dataset. Our implementation is based on PyTorch, using automatic mixed precision and trained on an NVIDIA A6000 GPU.

\begin{table*}[hbt!]
\centering
\resizebox{\textwidth}{!}{
\begin{tabular}{l c c c c c c c c c c} 
\toprule
\multirow{2}{*}{\centering \textbf{Model}} &
\multirow{2}{*}{\centering \textbf{\makecell{Piano\\Specific}}} &
\multirow{2}{*}{\textbf{\centering Item Modality}} &
\multicolumn{4}{c}{\textbf{In-domain} (199 tracks)} &
\multicolumn{4}{c}{\textbf{Out-of-domain} (88 tracks)} \\
\cmidrule(lr){4-7} \cmidrule(lr){8-11}
& & & R@1 & R@5 & R@10 & MedR↓ & R@1 & R@5 & R@10 & MedR↓ \\
\midrule
\textit{CLAP-Music}        & \xmark & Audio            & 0.00     & 7.38  & 9.85  & 54  & 5.13 & 20.51 & 41.03 & 15  \\
\textit{TTMR++}     & \xmark & Audio            & 1.47  & 6.40  & 12.31 & 45  & 5.13 & 15.38 & 28.21 & 16  \\
\textit{CLaMP}3\textsubscript{\textit{saas}}
      & \xmark & Audio            & 1.50  & 7.53  & 13.56  & 49  & 2.56 & 20.51 & 41.03 & 12  \\

\cmidrule(lr){1-11}
\textit{CLaMP2}     & \xmark & Symbolic         & 3.02 & 8.54  & 14.57 & 43  
  & 5.13 & 30.77 & 43.59 & 14  \\

\textit{CLaMP}$3^{c2}_{sa}$
      & \xmark & Symbolic         
      & 4.02  & 12.06 & 22.11 & 39  
      & 12.82 & 30.77 & 46.15 & 12  \\
\cmidrule(lr){1-11}

\textit{CLaMP}3\textsubscript{\textit{saas}}     & \xmark & Audio + Symbolic         
& 1.50 & 7.53  & 12.06 & 68  
& 2.56 & 33.33 & 46.15 & 13  \\

\textit{CLaMP}$3^{c2}_{sa}$ 
      & \xmark & Audio + Symbolic         
      & 2.51 & 10.55 & 17.58 & 47  
      & 7.69 & 28.20 & 43.58 & 13  \\
      
\textit{\textbf{PianoBind (Ours)}} & \textbf{\cmark} & \textbf{Audio + Symbolic} 
            & \textbf{10.55} & \textbf{35.67} & \textbf{52.76} & \textbf{10}
            & \textbf{15.38}  & \textbf{41.03} & \textbf{51.28} & \textbf{10} \\
\bottomrule
\end{tabular}
}
\caption{Performance comparisons between PianoBind and previous text-to-music retrieval models, conducted on the in-domain (PIAST-AT) and out-of-domain (EMOPIA-Caps) datasets.}
\vspace{-5pt}
\label{table:table2}
\end{table*}

\vspace{-5pt}
\section{Results}
\subsection{Comparison of Training Strategies}
\subsubsection{Multi-source training}
Table~\ref{table:table1} shows the performance of different training strategies across both in-domain (PIAST-AT) and out-of-domain (EMOPIA-Caps) test sets. We compare two multi-source learning approaches: combined training and pre-training followed by fine-tuning, each evaluated using two bimodal configurations (audio–text and symbolic–text) and one trimodal (audio–symbolic–text) configuration.

The results demonstrate that the pre-training and fine-tuning approach generally outperforms the combined training across modality configurations and metrics. In the in-domain setting, the pre-training and fine-tuning approach with trimodal integration achieves the best performance, reaching a Median Rank of 10, significantly surpassing the corresponding metrics for combined training. Similarly, in the out-of-domain context, it yields superior results with the same Median Rank. While a few isolated metrics—such as R@1 for audio in-domain and R@5 for symbolic out-of-domain—are marginally higher in the combined training setup, these exceptions do not contradict the overall trend. These findings underscore the challenges of small-scale annotated datasets. When dealing with limited high-quality annotations, the combined training approach exposes the model to data imbalance issues, where the larger but noisier dataset can potentially overpower the signal from the smaller expert-annotated data. In contrast, the sequential knowledge transfer approach—initially learning generalizable representations from broader data before adapting to specialized piano-specific contexts—enables the model to better leverage both data sources.

\subsubsection{Trimodal Integration}
Our results demonstrate the substantial performance gains achieved through trimodal integration compared to bimodal approaches. Across both multi-source training strategies, the trimodal model consistently outperforms both audio-text and symbolic-text configurations. The improvements are particularly pronounced in the pre-training and fine-tuning approach, where trimodal integration achieves a Median Rank of 10 compared to 12 for symbolic-only and 15 for audio-only. This performance advantage extends to higher R@10, with the trimodal approach achieving 52.76\%, significantly outperforming both symbolic and audio.

These findings strongly support our hypothesis that effective representation of piano music requires integrating multiple modalities. While both audio and symbolic representations capture valuable information—with symbolic representations slightly outperforming audio on in-domain retrieval—their combination in a trimodal framework yields representations that more comprehensively capture the semantic nuances of piano music. This suggests that audio and symbolic modalities provide complementary perspectives.

\subsection{Comparison with Existing Models}
Table~\ref{table:table2} presents a comparative analysis between PianoBind and existing text-to-music retrieval models. We compare against leading audio-based models—CLAP-Music, TTMR++, and CLaMP3${\textit{saas}}$ (optimized for audio)—as well as symbolic-based models, including CLaMP2 and CLaMP$3^{c2}_{sa}$ (optimized for symbolic). All models were trained on a large-scale of general-purpose datasets. The results demonstrate PianoBind's substantial performance advantage over general-purpose models. For in-domain retrieval, PianoBind achieves the lowest Median Rank of 10, significantly outperforming the best-performing general-purpose model, CLaMP$3^{c2}_{sa}$, which achieved a Median Rank of 39. This performance advantage extends to out-of-domain retrieval as well, where PianoBind maintains its lead with an R@10 of 51.28\% and Median Rank of 10, compared to the next best model, CLaMP$3^{c2}_{sa}$, with 46.15\% and 12, respectively.

Additionally, we also extended the models from CLaMP3, by implementing multimodal item embeddings through feature fusion—which is not present in the original work. However, even with this approach, CLaMP3 models with feature fusion underperform compared to their bi-modality results. This underperformance likely stems from the specialized nature of CLaMP3 variants, where each model was optimized for a specific modality.

\subsection{Comparative Analysis of Model Design Choices}
We conducted additional experiments to validate key model design choices. Specifically, we compared our averaged-loss training strategy in trimodal learning with the \textit{saas} (symbolic → audio → audio → symbolic) alignment strategy adopted in CLaMP3. As shown in Table~\ref{table:table3}, our averaged-loss training clearly outperforms both the original CLaMP3${\textit{saas}}$ model and our own reimplementation (Ours${\textit{\_saas}}$) in both in-domain and out-of-domain retrieval. These results demonstrate the advantage of jointly learning audio–text and MIDI–text embeddings, rather than aligning independently trained modalities through a staged alignment process. The consistent performance gains in Median Rank and R@10 further highlight the effectiveness of our unified training objective in capturing fine-grained semantic relationships in piano music.

\begin{table}[t]
\centering
\begin{tabular}{lcccc}
\toprule
\multirow{2}{*}{\textbf{Model}} 
& \multicolumn{2}{c}{\textbf{ID}} 
& \multicolumn{2}{c}{\textbf{OOD}} \\
\cmidrule(lr){2-3} \cmidrule(lr){4-5}
& R@10 & MedR↓ & R@10 & MedR↓ \\
\midrule
\textit{CLaMP3\_saas}           & 13.56 & 49 & 41.03 & 12 \\
\textit{Ours\_saas}       & 33.66 & 18 & 30.77 & 19 \\
\textbf{\textit{Ours\_LossAvg}}    & \textbf{52.76} & \textbf{10} & \textbf{52.76} & \textbf{10} \\
\bottomrule
\end{tabular}
\caption{Comparison between \textit{Saas} and Averaged Loss in trimodal learning in both in-domain (ID) and out-of-domain (OOD) evaluations.}
\vspace{-10pt}
\label{table:table3}
\end{table}

\vspace{-10pt}
\section{Conclusion}
In this paper, we introduced PianoBind, a multimodal joint embedding model designed for pop-piano music, integrating audio, symbolic, and textual modalities. Despite using substantially less training data than general-purpose models, PianoBind achieved strong retrieval performance. Our findings suggest that a sequential multi-source training strategy—pre-training on large-scale noisy data followed by fine-tuning on human-annotated examples—is more effective than training on the two sources simultaneously, particularly in low-resource settings. We also observed that integrating audio and symbolic modalities captures complementary semantic cues, and their joint use leads to more robust embeddings. Moreover, combining audio and symbolic embeddings at inference time improves retrieval performance, provided that the modalities are well-aligned through joint training. 

A limitation of our study is that our evaluation datasets, both in-domain and out-of-domain, are relatively small-scale, potentially restricting the generalizability of our findings. Moreover, our datasets primarily focused on pop-piano genres, lacking sufficient representation of classical and other diverse piano genres. Addressing these limitations by constructing larger-scale and more genre-diverse piano-text datasets remains an important direction for future work. This further highlights the need for comprehensive benchmarks to rigorously evaluate multimodal embedding models across varied piano music contexts.

\section{Acknowledgments}
This work has been supported by the collaboration with NCSOFT, Korea.

\bibliography{ISMIRtemplate}

\begin{thebibliography}{10}
\providecommand{\url}[1]{#1}
\csname url@samestyle\endcsname
\providecommand{\newblock}{\relax}
\providecommand{\bibinfo}[2]{#2}
\providecommand{\BIBentrySTDinterwordspacing}{\spaceskip=0pt\relax}
\providecommand{\BIBentryALTinterwordstretchfactor}{4}
\providecommand{\BIBentryALTinterwordspacing}{\spaceskip=\fontdimen2\font plus
\BIBentryALTinterwordstretchfactor\fontdimen3\font minus \fontdimen4\font\relax}
\providecommand{\BIBforeignlanguage}[2]{{%
\expandafter\ifx\csname l@#1\endcsname\relax
\typeout{** WARNING: IEEEtran.bst: No hyphenation pattern has been}%
\typeout{** loaded for the language `#1'. Using the pattern for}%
\typeout{** the default language instead.}%
\else
\language=\csname l@#1\endcsname
\fi
#2}}
\providecommand{\BIBdecl}{\relax}
\BIBdecl

\bibitem{popmusic}
Y.-S. Huang and Y.-H. Yang, ``Pop music transformer: Beat-based modeling and generation of expressive pop piano compositions,'' in \emph{Proceedings of the 28th ACM international conference on multimedia}, 2020, pp. 1180--1188.

\bibitem{compound}
W.-Y. Hsiao, J.-Y. Liu, Y.-C. Yeh, and Y.-H. Yang, ``Compound word transformer: Learning to compose full-song music over dynamic directed hypergraphs,'' in \emph{Proceedings of the AAAI Conference on Artificial Intelligence}, vol.~35, no.~1, 2021, pp. 178--186.

\bibitem{gen2}
S.-L. Wu and Y.-H. Yang, ``Compose \& embellish: Well-structured piano performance generation via a two-stage approach,'' in \emph{ICASSP 2023 - 2023 IEEE International Conference on Acoustics, Speech and Signal Processing (ICASSP)}, 2023, pp. 1--5.

\bibitem{gen3}
C.-P. Tan, H.~Ai, Y.-H. Chang, S.-H. Guan, and Y.-H. Yang, ``Picogen2: Piano cover generation with transfer learning approach and weakly aligned data,'' in \emph{Proceedings of the 25th International Society for Music Information Retrieval Conference (ISMIR)}, San Francisco, CA, United States, Nov. 2024.

\bibitem{amt1}
S.~Sigtia, E.~Benetos, and S.~Dixon, ``An end-to-end neural network for polyphonic piano music transcription,'' \emph{IEEE/ACM Transactions on Audio, Speech, and Language Processing}, vol.~24, no.~5, pp. 927--939, 2016.

\bibitem{amt2}
C.~Hawthorne, E.~Elsen, J.~Song, A.~Roberts, I.~Simon, C.~Raffel, J.~Engel, S.~Oore, and D.~Eck, ``Onsets and frames: Dual-objective piano transcription,'' \emph{arXiv preprint arXiv:1710.11153}, 2017.

\bibitem{amt3}
Q.~Kong, B.~Li, X.~Song, Y.~Wan, and Y.~Wang, ``High-resolution piano transcription with pedals by regressing onset and offset times,'' \emph{IEEE/ACM Transactions on Audio, Speech, and Language Processing}, vol.~29, pp. 3707--3717, 2021.

\bibitem{amt4}
\BIBentryALTinterwordspacing
T.~Kwon, D.~Jeong, and J.~Nam, ``Polyphonic piano transcription using autoregressive multi-state note model,'' in \emph{International Society for Music Information Retrieval Conference}, 2020. [Online]. Available: \url{https://api.semanticscholar.org/CorpusID:222125050}
\BIBentrySTDinterwordspacing

\bibitem{midibert}
Y.-H. Chou, I.~Chen, C.-J. Chang, J.~Ching, Y.-H. Yang \emph{et~al.}, ``{MidiBERT-Piano}: Large-scale pre-training for symbolic music understanding,'' \emph{arXiv preprint arXiv:2107.05223}, 2021.

\bibitem{pianobart}
X.~Liang, Z.~Zhao, W.~Zeng, Y.~He, F.~He, Y.~Wang, and C.~Gao, ``Pianobart: Symbolic piano music generation and understanding with large-scale pre-training,'' in \emph{2024 IEEE International Conference on Multimedia and Expo (ICME)}, 2024, pp. 1--6.

\bibitem{mulan}
Q.~Huang, A.~Jansen, J.~Lee, R.~Ganti, J.~Y. Li, and D.~P. Ellis, ``Mulan: A joint embedding of music audio and natural language,'' \emph{arXiv preprint arXiv:2208.12415}, 2022.

\bibitem{muscall}
I.~Manco, E.~Benetos, E.~Quinton, and G.~Fazekas, ``Contrastive audio-language learning for music,'' in \emph{Proceedings of the 23rd International Society for Music Information Retrieval Conference (ISMIR)}, 2022.

\bibitem{clap}
Y.~Wu*, K.~Chen*, T.~Zhang*, Y.~Hui*, T.~Berg-Kirkpatrick, and S.~Dubnov, ``Large-scale contrastive language-audio pretraining with feature fusion and keyword-to-caption augmentation,'' in \emph{IEEE International Conference on Acoustics, Speech and Signal Processing, ICASSP}, 2023.

\bibitem{ttmr}
S.~Doh, M.~Won, K.~Choi, and J.~Nam, ``Toward universal text-to-music retrieval,'' in \emph{ICASSP 2023-2023 IEEE International Conference on Acoustics, Speech and Signal Processing (ICASSP)}.\hskip 1em plus 0.5em minus 0.4em\relax IEEE, 2023, pp. 1--5.

\bibitem{ttmrpp}
S.~Doh, M.~Lee, D.~Jeong, and J.~Nam, ``Enriching music descriptions with a finetuned-llm and metadata for text-to-music retrieval,'' in \emph{ICASSP 2024 - 2024 IEEE International Conference on Acoustics, Speech and Signal Processing (ICASSP)}, 2024, pp. 826--830.

\bibitem{clamp}
S.~Wu, D.~Yu, X.~Tan, and M.~Sun, ``Clamp: Contrastive language-music pre-training for cross-modal symbolic music information retrieval,'' \emph{arXiv preprint arXiv:2304.11029}, 2023.

\bibitem{clamp2}
S.~Wu, Y.~Wang, R.~Yuan, Z.~Guo, X.~Tan, G.~Zhang, M.~Zhou, J.~Chen, X.~Mu, Y.~Gao \emph{et~al.}, ``Clamp 2: Multimodal music information retrieval across 101 languages using large language models,'' \emph{arXiv preprint arXiv:2410.13267}, 2024.

\bibitem{clamp3}
\BIBentryALTinterwordspacing
S.~Wu, Z.~Guo, R.~Yuan, J.~Jiang, S.~Doh, G.~Xia, J.~Nam, X.~Li, F.~Yu, and M.~Sun, ``Clamp 3: Universal music information retrieval across unaligned modalities and unseen languages,'' 2025. [Online]. Available: \url{https://arxiv.org/abs/2502.10362}
\BIBentrySTDinterwordspacing

\bibitem{piast}
H.~Bang, E.~Choi, M.~Finch, S.~Doh, S.~Lee, G.-H. Lee, and J.~Nam, ``{PIAST}: A multimodal piano dataset with audio, symbolic and text,'' in \emph{Proceedings of the 3rd Workshop on NLP for Music and Audio (NLP4MusA)}, Nov. 2024, pp. 5--10.

\bibitem{pianotree}
Z.~Wang, Y.~Zhang, Y.~Zhang, J.~Jiang, R.~Yang, J.~Zhao, and G.~Xia, ``Pianotree vae: Structured representation learning for polyphonic music,'' \emph{arXiv preprint arXiv:2008.07118}, 2020.

\bibitem{polydis}
Z.~Wang, D.~Wang, Y.~Zhang, and G.~Xia, ``Learning interpretable representation for controllable polyphonic music generation,'' \emph{Proceedings of the 23rd International Society for Music Information Retrieval Conference (ISMIR)}, 2020.

\bibitem{collagenet}
A.~Wuerkaixi, C.~Benetatos, Z.~Duan, and C.~Zhang, ``Collagenet: Fusing arbitrary melody and accompaniment into a coherent song,'' \emph{International Society for Music Information Retrieval}, 2022.

\bibitem{emopia}
H.-T. Hung, J.~Ching, S.~Doh, N.~Kim, J.~Nam, and Y.-H. Yang, ``{EMOPIA}: A multi-modal pop piano dataset for emotion recognition and emotion-based music generation,'' in \emph{Proceedings of 22th International Conference on Music Information Retrieval (ISMIR)}, 2021.

\bibitem{resnet}
K.~He, Y.~Wang, and J.~Hopcroft, ``A powerful generative model using random weights for the deep image representation,'' \emph{Advances in Neural Information Processing Systems}, vol.~29, 2016.

\bibitem{roberta}
\BIBentryALTinterwordspacing
Y.~Liu, M.~Ott, N.~Goyal, J.~Du, M.~Joshi, D.~Chen, O.~Levy, M.~Lewis, L.~Zettlemoyer, and V.~Stoyanov, ``Roberta: {A} robustly optimized {BERT} pretraining approach,'' \emph{CoRR}, vol. abs/1907.11692, 2019. [Online]. Available: \url{http://arxiv.org/abs/1907.11692}
\BIBentrySTDinterwordspacing

\bibitem{infoNCE}
A.~v.~d. Oord, Y.~Li, and O.~Vinyals, ``Representation learning with contrastive predictive coding,'' \emph{arXiv preprint arXiv:1807.03748}, 2018.

\bibitem{mt3}
J.~Gardner, I.~Simon, E.~Manilow, C.~Hawthorne, and J.~Engel, ``Mt3: Multi-task multitrack music transcription,'' \emph{arXiv preprint arXiv:2111.03017}, 2021.

\bibitem{joint}
Y.~Wu, K.~Chen, T.~Zhang, Y.~Hui, T.~Berg-Kirkpatrick, and S.~Dubnov, ``Large-scale contrastive language-audio pretraining with feature fusion and keyword-to-caption augmentation,'' in \emph{ICASSP 2023-2023 IEEE International Conference on Acoustics, Speech and Signal Processing (ICASSP)}.\hskip 1em plus 0.5em minus 0.4em\relax IEEE, 2023, pp. 1--5.

\bibitem{lpcaps}
S.~Doh, K.~Choi, J.~Lee, and J.~Nam, ``Lp-musiccaps: Llm-based pseudo music captioning,'' in \emph{ISMIR}, 2023.

\bibitem{audioclip}
A.~Guzhov, F.~Raue, J.~Hees, and A.~Dengel, ``Audioclip: Extending clip to image, text and audio,'' in \emph{ICASSP 2022-2022 IEEE International Conference on Acoustics, Speech and Signal Processing (ICASSP)}.\hskip 1em plus 0.5em minus 0.4em\relax IEEE, 2022, pp. 976--980.

\bibitem{gpt}
OpenAI, ``Chatgpt-4o (gpt-4 omni),'' \url{https://openai.com/index/gpt-4o}, 2024, accessed: 2025-03-26.

\end{thebibliography}

\end{document}